\documentstyle[prd,aps,floats]{revtex}

\begin{document}
\draft

\input epsf 
\renewcommand{\topfraction}{0.8}
\def\ds{\displaystyle}

\title{Inflatonic solitons in running mass inflation}
\author{Kari Enqvist~$^{a,b}$, Shinta Kasuya~$^b$,
        and Anupam Mazumdar~$^c$}
\address{$^a$ Department of Physical Sciences, P. O. Box 64,
              FIN-00014, University of Helsinki, Finland.\\
         $^b$ Helsinki Institute of Physics, P. O. Box 64,
              FIN-00014, University of Helsinki, Finland.\\
         $^c$ The Abdus Salam International Centre for Theoretical
              Physics, I-34100, Trieste, Italy.}
\date{June 26, 2002}

\maketitle

\begin{abstract}
The inflaton condensate associated with a global symmetry can fragment
into quasistable $Q$ balls, provided the inflaton oscillations give
rise to an effective equation of state with negative pressure. We
study chaotic inflation with a running inflaton mass and show that,
depending on the sign of the radiative mass correction, the process of
fragmentation into inflatonic $Q$ balls can actually take place 
even though no net charge exists. If the main decay channel of the $Q$
ball is to fermions, the universe will be reheated slowly via surface
evaporation. 
\end{abstract}

\pacs{PACS numbers: 98.80.Cq, 11.27.+d, 11.30.Fs
      \hspace{21.5mm} HIP-2002-28/TH, hep-ph/0206272}


\section{Introduction}

Reheating of the universe, in which  the inflaton decay product
provide a thermal bath of radiation, is one of the cornerstones of any
inflationary model. At the end of inflation the coherent inflaton
condensate may decay either perturbatively \cite{pert}, one quantum at
the time, or non-perturbatively \cite{traschen} via a parametric
resonance in a collective process denoted as preheating. Usually in
the literature reheating has been considered to be a volume effect so
that it tends to be relatively fast, and as a consequence, the reheat
temperature is relatively high. In supersymmetric models this has
turned out to be somewhat problematic because a large reheat
temperature generates a large gravitino abundance which is in
conflict with the successful big bang nucleosynthesis \cite{gravitino}. 

However, recently we have pointed out \cite{EKM} a novel possibility
in which reheating could be realized as a surface effect, noting that 
the evaporation rate from a surface is always suppressed by the area
of the evaporating surface. Therefore the larger the surface, the
smaller the reheat temperature of the universe. Such a situation can
be achieved if the zero-mode inflaton condensate does not remain
spatially coherent but breaks down into quasi-stable lumps. This can
happen if the inflaton field carries a global $U(1)$ charge which
protects the lumps; for instance, the inflaton could be a complex
rather than a real scalar field. The ground state of such a condensate
lump would be a non-topological soliton called $Q$ ball
\cite{coleman}, the properties of which depend on the exact form of
the scalar potential. A general requirement for its existence is that
the scalar potential should grow slower than $\vert \Phi\vert^2$.

Inflaton condensate fragmentation has recently been discussed in the 
context of hybrid inflation models both numerically \cite{copeland}
and analytically \cite{john}. However, in these cases the shape of the
potential does not render the condensate lumps quasi-stable and their
appearance is merely a part of the complicated preheating dynamics. In
fact, there they appear to make reheating even more effective rather
than slowing it down \cite{john}. 

$Q$ ball formation has been considered in detail in the context of the
MSSM flat direction condensates (Affleck-Dine condensates \cite{AD}),
and can be either completely stable \cite{DKS,KuSh,KK1,KK4,KK3} if
SUSY is broken by gauge mediation, or quasi-stable \cite{empl,em,KK2}
if SUSY is broken by gravity mediation. In the Affleck-Dine case the
condensate is charged up by a helical motion in the field space, and
fragmentation takes place when the fluctuations in the real and the
imaginary directions grow to become of order one. 

In this paper we consider the simple case of the chaotic inflation but
with a complex inflaton field (which is rather more natural) and a running 
inflaton mass, such as might be generated radiatively by the Yukawa 
and/or gauge couplings of the inflaton to other fields. We demonstrate 
numerically that at the end of inflation the inflaton condensate fragments 
into lumps and forms $Q$ balls. The initial fragmentation occurs because 
of the negative pressure the condensate feels during the inflaton
oscillations. We show that the fluctuations in the imaginary direction
grow because of the mode-mode coupling between the real and the
imaginary parts and thus provide the motion in field space necessary
for solitogenesis. 

The paper is organized as follows. In Section II we describe
$Q$ balls in a running mass potential and discuss how the fluctuations
grow for a certain running mass inflaton potential in Section III. In
Section IV we study $Q$-ball formation due to mode-mode couplings by
lattice simulations. Section V is devoted to the implications of
$Q$-ball formation and a discussion on the reheat temperature. In
Section VI we conclude with a brief discussion highlighting the
results and describing future prospects.

\section{$Q$ balls in the running mass inflaton potential}

If the inflaton has couplings to other fields, then, in general, the
inflaton mass should receive radiative corrections \cite{LyRi}. 
The simplest chaotic inflation potential with a running inflaton
mass could then be written as
\begin{equation}
    \label{qpot}
    V = m^2 |\Phi|^2
    \left[ 1 + K\log\left(\frac{|\Phi|^2}{M^2}\right)\right ]\,,
\end{equation}
where the coefficient $K$ could be negative or positive, and $m$ is
the bare mass of the inflaton. The logarithmic correction to
the mass of the inflaton is something one would expect to arise
because of the possible Yukawa and/or gauge couplings to other fields.

Since we assume that $|K|\ll 1$, the second term in the bracket does 
not affect the inflaton dynamics or generation of perturbations 
much so that with a running mass the universe would experience the
same chaotic inflationary stage as with the usual  $m^2|\Phi|^2$
potential. However, as we shall see, the values $K<0$ lead to 
interesting cosmological consequences. Though it is not pertinent, we
note that the potential Eq.~(\ref{qpot}) can be generated in a
supersymmetric theory if the inflaton has a gauge coupling
\cite{em,EKM}. This is due to the fact that the gaugino loops
contribute as a negative correction which dominates. In this case the
value of $K$ is given by 
\begin{equation}
    K \sim -\frac{\alpha}{8\pi}
        \frac{m_{1/2}^2}{m_{\tilde{\ell}}^2}\,,
\end{equation}
where $m_{1/2}$ is the gaugino mass and $m_{\tilde{\ell}}$ denotes the
mass of slepton. It is also possible to obtain the  potential
Eq.~(\ref{qpot}) in a non-supersymmetric (or in broken SUSY) theory,
provided the fermions live in a larger representation than the
bosons. In this latter situation the value of $K$ is determined by the
Yukawa coupling 
\cite{EKM}
\begin{equation}
    K =-C\frac{h^2}{16\pi^2}\,.
\end{equation}
Here $C$ is the total number of fermionic loops, and $h$ is the Yukawa
coupling. As long as $|K| \ll 1$, the dominant contribution to the
potential comes from $m^2|\Phi|^2$ term during inflation, and
inflationary slow roll conditions are satisfied as in the case of the
standard chaotic model. The generation of the amplitude of the density
perturbations are constrained by the COBE normalization, which implies
\begin{equation}
    \frac{\delta \rho}{\rho} \sim \frac{m}{M_{\rm P}}\sim 10^{-5}\,,
\end{equation}
which results in $m\sim 10^{13}$~GeV.

The most important virtue of the inflationary potential
Eq.~(\ref{qpot}) is that it respects a global $U(1)$ symmetry, and
for a negative $K$, the potential is shallower than $m^2|\Phi|^2$. It
has been well known that such a potential admits a $Q$-ball solution
\cite{em}. This means that the energy of a 
$Q$-ball configuration is less than a collection of free scalars
carrying an equivalent charge. The field configuration is given by
$\Phi(x,t)= (e^{i\omega t}\phi(x))/\sqrt{2}$ while its energy and
charge read
\begin{eqnarray}
    \label{energy}
    E & = & \int d^3x \left[ \frac{1}{2}\omega^2\phi^2
      + \frac{1}{2} (\nabla \phi)^2 + V(\phi) \right]\,, \nonumber \\
    Q & = & \omega\int \phi^2 d^3x\,.
\end{eqnarray}
Here we assume that the potential $V(\phi)$ has a global minimum at
the origin and  is invariant under a global $U(1)$-transformation.

Depending on the slope of the potential it is possible to have
an energetically favorable state in the form of $Q$ balls. The energy
of the $Q$ ball is $E = \mu Q$, where $\mu$ lies in the range 
$\mu_0  \leq \mu < m_0$, and $\mu_0$ and $m_0$ are given by
\begin{equation}
    \label{mu}
    \mu_0 = {\rm min}\left(\sqrt{\frac{2 V(\phi)}{\phi^2}}\right),
    \quad
    m_0 = \sqrt{V''(0)}.
\end{equation}
The $Q$ ball is stable against decay into other particles if
$\mu q < m_{decay}$, where $m_{decay}$ is the smallest mass of the
particles which shares the same $U(1)$ charge as the complex field
$\Phi$, and $q$ is the charge carried by the particle. In this case, a
$Q$ ball exists forever. On the other hand, if the stability condition
does not hold, $Q$ balls will eventually decay into lighter
particles. However, if the (final) decay products are pairs of
fermions, the decay should proceed by the evaporation of charges
through the surface of $Q$ balls because of the Pauli blocking
\cite{Cohen}. Thus, inflatonic $Q$ balls with a large 
enough charge could live very long, in which case
reheating may take place very slowly, as we shall discuss.

The potential of the form Eq.~(\ref{qpot}) generically forms a
thick-wall $Q$ balls with a Gaussian profile \cite{em}
\begin{equation}
    \phi(r)\propto \exp(-|K|m^2r^2)\,,
\end{equation}
so that the size of such a $Q$ ball can be estimated by
\begin{equation}
    R \simeq |K|^{-1/2}m^{-1}\,.
\end{equation}
The energy of a $Q$ ball can be approximately given by
\begin{equation}
    E \simeq m Q\,,
\end{equation}
and the total charge accumulated by a $Q$ ball is given by
\begin{equation}
    Q \simeq \omega \phi_0^2 V \sim \frac{4\pi}{3}m \phi_0^2 R^3\,,
\end{equation}
where $\phi_0$ is the field value at which the $Q$ ball forms, and 
$V\sim 4\pi R^3/3$ is the volume of the $Q$ ball.

\section{Amplification of fluctuations}

In the context of Affleck-Dine baryogenesis, the homogeneous mode is
rotating around the origin in the potential. In such a case
fluctuations tend to grow when on the average there is a negative 
pressure, which can be achieved in the potential Eq.~(\ref{qpot}) when
$K<0$ \cite{turnerpressure,johnpressure}. Expanding the potential in
Eq.~(\ref{qpot}) one finds 
\begin{equation}
    \label{pot0}
    V(\phi) \simeq \frac{1}{2}m^2\phi^2
    \left(\frac{\phi^2}{2M^2}\right)^K \propto \phi^{2+2K}\,.
\end{equation}
where we assume $|K|\ll 1$, since this term is actually a correction
to a tree level potential. The equation of state for a field rotating
in such a potential is then given by
\footnote{
Because in the Affleck-Dine case the exact motion of the field is
helical and hence strictly speaking non-periodic, 
the equation of state actually oscillates \cite{asko}.}
\begin{equation}
    \label{state}
    p\simeq \frac{K}{2+K} \rho \simeq -\frac{|K|}{2}\rho \,,
\end{equation}
where $p$ and $\rho$ is a pressure and energy density of the scalar
field, respectively. Therefore, negative value of $K$ is crucial for
negative pressure, which is the core of our discussion. Notice the
difference between this case and the usual chaotic potential
$V=m^2|\Phi|^2$, where on the average the oscillating inflaton
condensate acts as a pressureless fluid.

When the homogeneous mode is rotating in the potential, it is
appropriate to write a complex field as
$\Phi = (\phi e^{i \theta})/\sqrt{2}$,
and decompose it into a homogeneous part and fluctuations:
$\phi \rightarrow \phi +\delta\phi$ and
$\theta \rightarrow \theta + \delta\theta$. The equations of
motion of the field read as \cite{KuSh,em}
\begin{eqnarray}
    \ddot{\phi} + 3H\dot{\phi} - \dot{\theta}^2\phi
             + V'(\phi) & = & 0, \\
    \label{theta-eq}
    \phi\ddot{\theta} + 3H\phi\dot{\theta}
      + 2\dot{\phi}\dot{\theta} & = & 0,
\end{eqnarray}
for the homogeneous mode, and
\begin{eqnarray}
    \label{eom-fl-1}
    \delta\ddot{\phi} + 3H\delta\dot{\phi}
       - 2\dot{\theta}\phi\delta\dot{\theta}
       - \dot{\theta}^2\delta\phi  -\frac{\nabla^2}{a^2}\delta\phi
       + V''(\phi)\delta\phi & = & 0, \\
    \label{eom-fl-2}
    \phi\delta\ddot{\theta}
       + 3H\phi\delta\dot{\theta}
       + 2(\dot{\phi}\delta\dot{\theta}
             + \dot{\theta}\delta\dot{\phi})
       -2\frac{\dot{\phi}}{\phi}\dot{\theta}\delta\phi
       -\phi\frac{\nabla^2}{a^2}\delta\theta & = & 0,
\end{eqnarray}
for the fluctuations,
and
\begin{eqnarray}
    V'(\phi) & = & m^2 \phi \left[ 1 + K +
       K\log\left(\frac{\phi^2}{2M^2}\right) \right], \\
    V''(\phi) & = & m^2 \left[ 1 + 3K +
       K\log\left(\frac{\phi^2}{2M^2}\right) \right].
\end{eqnarray}
Equation (\ref{theta-eq}) represents the conservation of the charge
(or number) within the physical volume: $\dot{\theta}\phi^2a^3=const.$
Since the energy density of the scalar field dominates the universe,
the homogeneous part of the field evolves as
\begin{eqnarray}
    \phi(t) & \simeq &
       \left( \frac{a(t)}{a_0} \right)^{-3/(2+K)} \phi_0, \\
    \dot{\theta}^2(t) &\simeq &
       \left( \frac{a(t)}{a_0} \right)^{-6K/(2+K)} m^2.
\end{eqnarray}

We are now going to find the most amplified mode. To this end, we take
the solutions in the form
\begin{eqnarray}
    \delta\phi & = & \left( \frac{a(t)}{a_0} \right)^{-3/(2+K)}
      \delta\phi_0 e^{\alpha(t)+ikx}, \\
    \delta\theta & = & \left( \frac{a(t)}{a_0} \right)^{-3K/(2+K)}
      \delta\theta_0 e^{\alpha(t)+ikx}.
\end{eqnarray}
If $\dot{\alpha}$ is real and positive, these fluctuations grow
exponentially, and go nonlinear to form $Q$ balls. Putting these forms
into Eqs.(\ref{eom-fl-1}) and (\ref{eom-fl-2}), we get the following
condition for non-trivial $\delta\phi_0$ and $\delta\theta_0$,
\begin{equation}
  \label{det}
   \left|
      \begin{array}{cc}
          \ds{\frac{3K}{2+K}H\dot{\alpha}+\ddot{\alpha}
          +\dot{\alpha}^2+\frac{k^2}{a^2}+2m^2Ka^{-6K/(2+K)}}
          & \ds{-2ma^{-6K/(2+K)}\phi_0
              \left(-\frac{3K}{2+K}H+\dot{\alpha}\right)} \\[3mm]
          \ds{\frac{2m\dot{\alpha}}{\phi_0}}
          & \ds{\ddot{\alpha}+\dot{\alpha}^2+\frac{k^2}{a^2}
            +\frac{3K}{2+K}\left[ (4-3K)H^2 -\frac{\ddot{a}}{a}
              -H\dot{\alpha}\right]}
    \end{array}
    \right| = 0,
\end{equation}
where we set $a_0=1$.

Let us neglect the cosmological expansion and assume
$\ddot{\alpha} \ll \dot{\alpha}^2$ for simplicity. Then,
Eq.(\ref{det}) will be simplified as
\begin{equation}
  \label{det2}
   \left|
      \begin{array}{ccc}
          \ds{\dot{\alpha}^2+\frac{k^2}{a^2}+2m^2K}
          & &\ds{-2m\phi_0\dot{\alpha}} \\[2mm]
          \ds{\frac{2m\dot{\alpha}}{\phi_0}}
          & & \ds{\dot{\alpha}^2+\frac{k^2}{a^2}}
    \end{array}
    \right| = 0.
\end{equation}
In order for $\dot{\alpha}$ to be real and positive, we must have
\begin{equation}
    \frac{k^2}{a^2}\left( \frac{k^2}{a^2}+2m^2K \right) < 0.
\end{equation}
As we are considering negative $K$, an instability band exists. This is 
because the rotating inflaton field in the potential flatter than 
$\phi^2$ has a negative pressure. Thus the instability band
should be found in the range
\begin{equation}
    \label{k-band}
    0 < \frac{k^2}{a^2} < \frac{k_{max}^2}{a^2} \equiv  2m^2|K|.
\end{equation}
We can easily derive that the most amplified mode lies at about the
center of the band: $(k_{most}/a)^2 \simeq m^2|K|(1-|K|/4)$, and the
maximum growth rate is $\dot{\alpha}_{most} =|K|m_{3/2}/2$.
\footnote{
The coefficients of these results are slightly different from the
previous estimations \cite{KK2,KK3} because these were based on the
approximation $\dot{\theta}^2 \approx m^2$. Replacing this with the
more accurate 
$\dot{\theta}^2 = V'(\phi)/\phi=m^2[1+K+K\log(\phi^2/2M^2)]$,
and still assuming a  matter-dominated universe, one finds the
same estimates as in the present paper.}

After inflation the inflaton field is just oscillating around the
origin of its potential. It is thus more appropriate to
decompose the inflaton field into its real and imaginary parts for
analyzing the dynamics when the motion is non-circular.
For numerical calculations it is convenient to take
all the variables to be dimensionless, so we normalize as
$\varphi=\phi/m$, $\tilde{k}=k/m$, $\tau = mt$, and $\xi = mx$.
Writing $\varphi = (\varphi_1 + i \varphi_2)/\sqrt{2}$, we get the
equations for the homogeneous mode as
\begin{equation}
    \varphi_i''+ 3h\varphi_i' + \left[ 1 + K + K\log \left(
       \frac{\varphi_1^2+\varphi_2^2}{2M^2} \right) \right]\varphi_i
    = 0,
\end{equation}
where $h=H/m$, $i=1,2$, and the prime denotes the derivative with
respect to $\tau$, and, for the fluctuations,
\begin{equation}
    \left[\frac{d^2}{d\tau^2} + 3h\frac{d}{d\tau}
      + \frac{\tilde{k}^2}{a^2} + V_{ij}\right]
    \left(\begin{array}{c}
            \ds{\delta\varphi_1} \\
            \ds{\delta\varphi_2}
          \end{array} \right) = 0,
\end{equation}
where $V_{ij}$ denotes the second derivative with respect to
$\varphi_i$ and $\varphi_j$, and explicitly written as
\begin{eqnarray}
    & &
    V_{ii} = 1 + K + K\log \left(
              \frac{\varphi_1^2+\varphi_2^2}{2M^2} \right)
             + 2K \frac{\varphi_i^2}{\varphi_1^2+\varphi_2^2},
    \nonumber \\
    & &
    V_{12} = V_{21} = 2K \frac{\varphi_1\varphi_2}
    {\varphi_1^2+\varphi_2^2},
\end{eqnarray}

\begin{figure}[t!]
\centering
\hspace*{-7mm}
\leavevmode\epsfysize=6.5cm \epsfbox{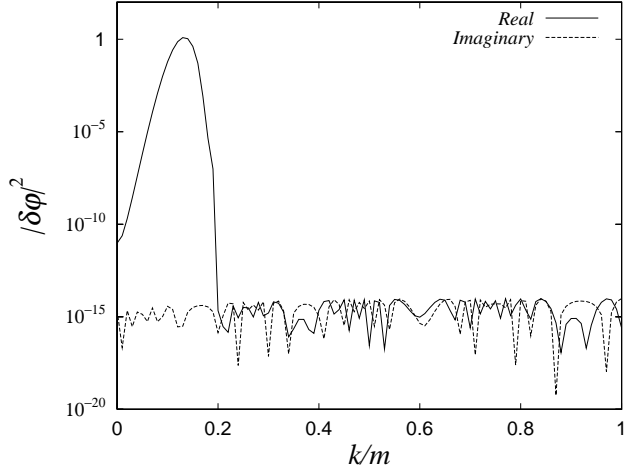}\\[2mm]
\caption[fig1]{\label{mode-osc}
Instability bands when the homogeneous mode is just oscillating along
the real direction in the potential (\ref{qpot}). Solid and dashed
lines denote the fluctuations in the real and imaginary directions,
respectively. Here we take $K=-0.02$.}
\end{figure}

\begin{figure}[t!]
\centering
\hspace*{-7mm}
\leavevmode\epsfysize=6.5cm \epsfbox{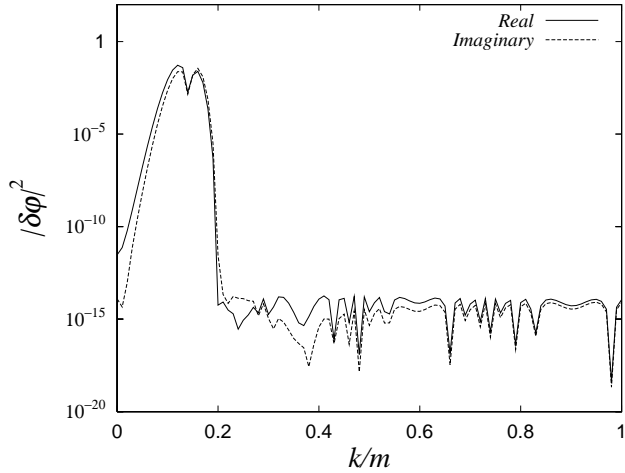}\\[2mm]
\caption[fig2]{\label{fig2}
Instability band when the homogeneous mode is rotating circularly in
the potential (\ref{qpot}). Solid and dashed lines denote the
fluctuations in the real and imaginary directions, respectively.
Here we take $K=-0.02$.}
\end{figure}

In particular, when the homogeneous part is just oscillating along the
radial (i.e., real) direction, we have
\begin{equation}
    \delta\varphi_i'' + 3h \delta\varphi_i' 
      + \left[ \frac{\tilde{k}^2}{a^2} + 1 + \beta_i K
      + K\log \left( \frac{\varphi_1^2+\varphi_2^2}{2M^2} \right)
          \right]\delta\varphi_i = 0,
\end{equation}
where $\beta_1=3$ and $\beta_2=1$. At first glance, we could not see
any crucial difference between evolutions of real and imaginary parts
of fluctuations. As we see in Fig.~\ref{mode-osc}, however, the
imaginary part does not grow at all, while considerable growth can be
seen in the real part. (Here we have neglected cosmological
expansion for simplicity.) Notice that both real and imaginary parts
of the fluctuations develop when the homogeneous part is rotating in
the potential.\footnote{
When the field is rotating in the potential, fluctuations do not grow
for positive $K$. However, the imaginary part of the fluctuations
actually grows when the homogeneous field is just oscillating. The
growth rate is much larger than that for the negative $K$, but,
needless to say, $Q$-ball formation never occurs, since a $Q$-ball
solution does not exist for positive $K$.}
(See Fig.~\ref{fig2}.) In either case, the instability band coincides
with the analytical estimate (\ref{k-band}).

One might thus be tempted to conclude that there is no $Q$-ball
formation, but, as we will see in the next Section, mode-mode (self)
couplings play a crucial role for developing the imaginary part of the
fluctuations, resulting in $Q$ and anti-$Q$ balls to form.

\section{Mode-mode couplings}

It is not apparent that the fluctuations in the imaginary direction
grows on the top of the pure oscillating homogeneous condensate in the
real direction. As mentioned in the previous Section, however,
self-couplings of the inflaton field are expected to create
fluctuations in the imaginary direction as well. This can be best seen
in lattice simulations, which we have done in one, two, and three
spatial dimensions. Since, all the cases are more or less the same in
the qualitative sense, and the resolution of the three dimensional
lattices is not so good, we have grounds for believing that the
two-dimensional results presented here will also capture the 
qualitative behavior of the full three-dimensional case. In the
context of Affleck-Dine baryogenesis, lattice simulations have been
done previously \cite{KK1,KK2,KK3,MuVi,EnJoMuVi}, but it is usually
assumed that the homogeneous condensate is rotating, not just
oscillating in the potential. In addition, here the universe is
dominated by the inflaton field in contrast to the Affleck-Dine case,
in which a matter-dominated universe is usually assumed. We also note
that $Q$-ball formation in the pure oscillation of the homogeneous
condensate has been numerically verified previously in the context of
Affleck-Dine mechanism \cite{Kasuya1}, and discussed in the context of
quintessence and oscillating inflation \cite{Kasuya2}.

After inflation the inflaton field is almost homogeneous beyond the
horizon, and its motion in phase direction has completely
frozen. However, there exist small variations, which are 
the seeds for the large scale structures in the universe.
We can thus pose the initial condition at each lattice site as
follows:
\begin{eqnarray}
    & & \phi_1(0)=\phi_0 + \Delta_1, \qquad
        \dot{\phi}_1(0)=\dot{\phi}_0 + \Delta_2 \nonumber \\
    & & \phi_2(0)= \Delta_3, \qquad \dot{\phi}_2(0)=\Delta_4,
\end{eqnarray}
where $\phi_0=M_P/\sqrt{12\pi}$, $\dot{\phi}_0=-mM_P/\sqrt{12\pi}$,
and $\Delta$'s are small numbers which represent small fluctuations.
The equations to be integrated are
\begin{equation}
    \varphi_i'' + 3h\varphi_i'
          - \frac{1}{a^2}\nabla^2\varphi_i
          + \varphi_i \left[ 1+K+K\log\left\{
              \frac{ \tilde{m}^2 (\varphi_1^2+\varphi_2^2)}{2}
                 \right\} \right] = 0,
\end{equation}
where $i=1$ and 2, which denote real and imaginary parts of the field,
respectively, and $\tilde{m}=m/M$. Here we will mainly discuss results
from two-dimensional lattices with the box size $N=2048$ and the
lattice spacing $\Delta \xi=0.004$, and $K=-0.1$, and show qualitative
three-dimensional results with the box size $N=128$, the lattice
spacing $\Delta \xi=0.01$, and $K=-0.1$ as well.

\begin{figure}[t!]
\centering
\hspace*{-7mm}
\leavevmode\epsfysize=6.5cm \epsfbox{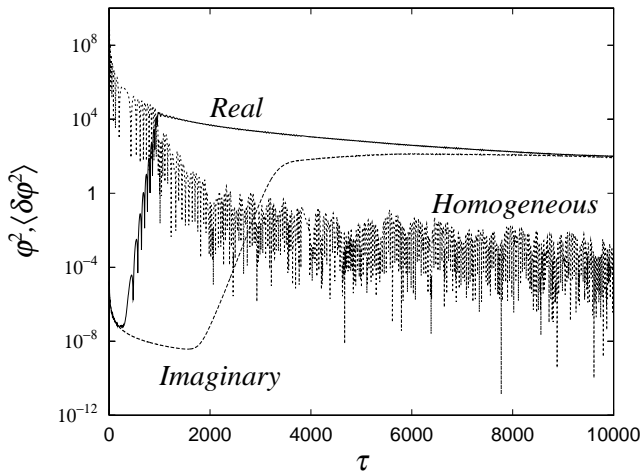}\\[2mm]
\caption[fig3]{\label{evolution}
Evolutions of the amplitudes of homogeneous mode, and fluctuations in
the real and imaginary directions.}
\end{figure}

Fluctuations in the real direction grow very fast, and their amplitudes
become of the same order as that of homogeneous mode at 
$\tau\sim 1000$. It is depicted in Fig.~\ref{evolution}, where we show
the time evolution of both real and imaginary parts of the
fluctuations. Fluctuations in the imaginary direction, however, have
not yet grown at that time, and it starts growing later.

\begin{figure}[t!]
\centering
\hspace*{-7mm}
\leavevmode\epsfysize=6.5cm \epsfbox{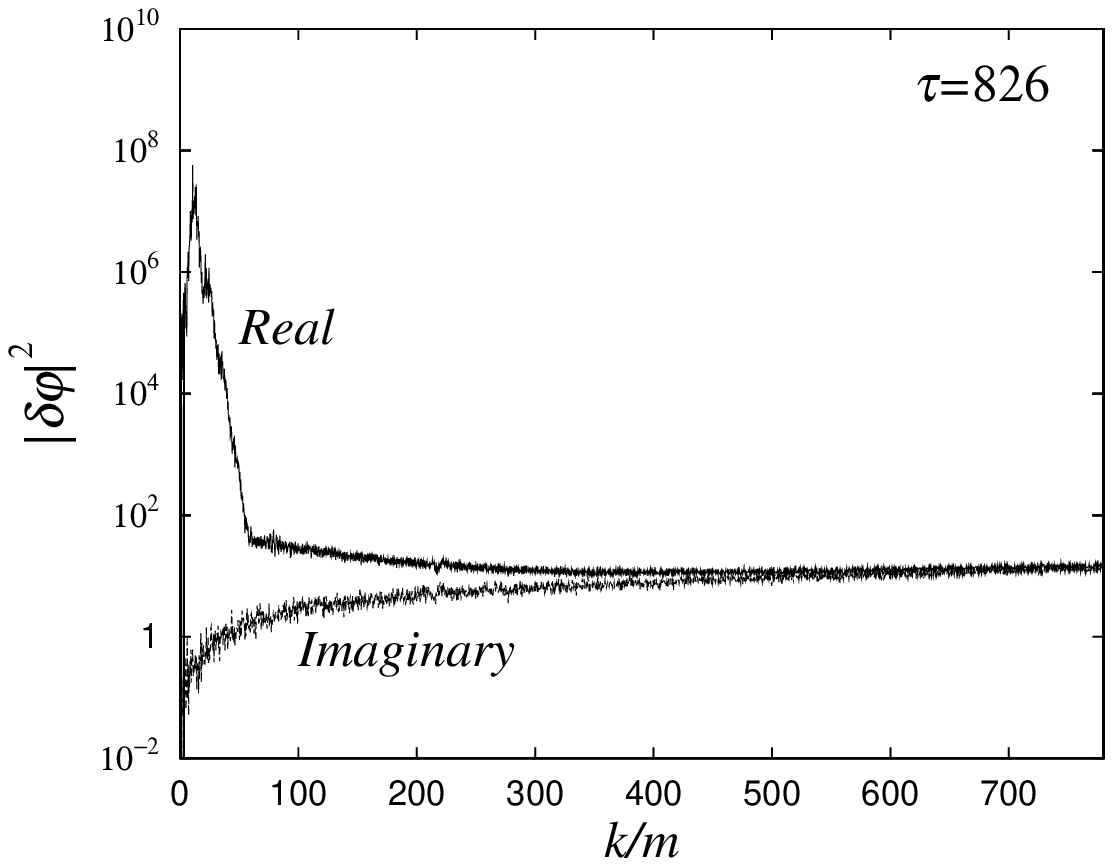}\\[2mm]
\caption[fig3]{\label{sp826}
Power spectra of fluctuations in the real and imaginary directions
just before the real part fully develops.}
\end{figure}

\begin{figure}[t!]
\centering
\hspace*{-7mm}
\leavevmode\epsfysize=6.5cm \epsfbox{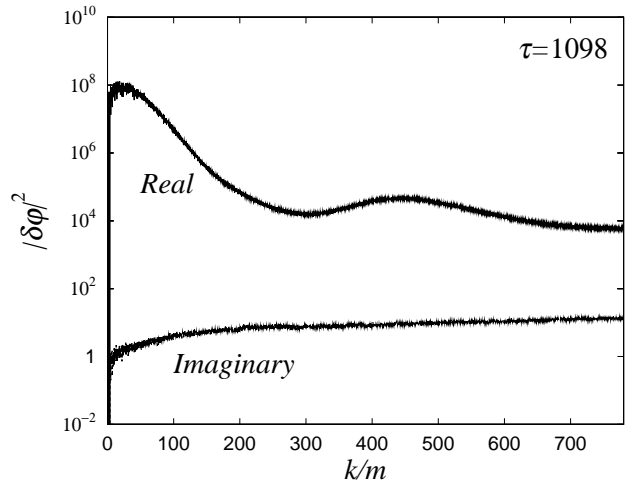}\\[2mm]
\caption[fig3]{\label{sp1098}
Power spectra of fluctuations in the real and imaginary directions
just after the real part fully developed.}
\end{figure}

In Fig.~\ref{sp826}, we plot the spectra of real and imaginary parts
of the fluctuations at $\tau=826$, just before the former fully
develops to the size of the background field value. In fact, the
instability band lies in the right range  
$k \lesssim a(\tau)|K|^{1/2} m$, and the most amplified mode coincides
with the anticipated $Q$-ball size. After the real part of the
fluctuations fully developed, other modes outside the instability band
also grow due to mode-mode couplings, as can be seen in
Fig.~\ref{sp1098}, which shows the spectra of the fluctuations at
$\tau=1098$, just after the real part has fully developed. Although
the spectrum becomes wider and smoother, the position of the most
amplified mode, which corresponds the $Q$-ball size, stays still. 

\begin{figure}[t!]
\centering
\hspace*{-7mm}
\leavevmode\epsfysize=6.5cm \epsfbox{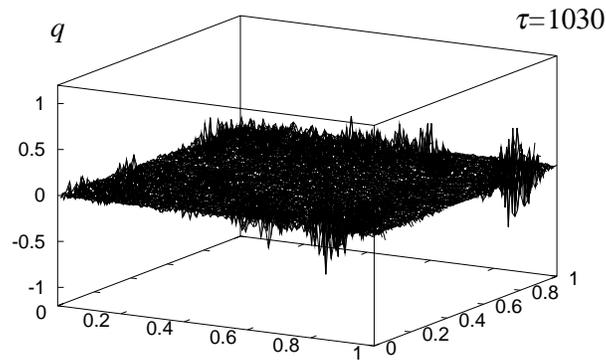}\\[2mm]
\caption[fig3]{\label{q_1030}
Charge density of the field. Here we show only a small portion of the 
entire lattice of $8.192\times 8.192$.}
\end{figure}

\begin{figure}[t!]
\centering
\hspace*{-7mm}
\leavevmode\epsfysize=6.5cm \epsfbox{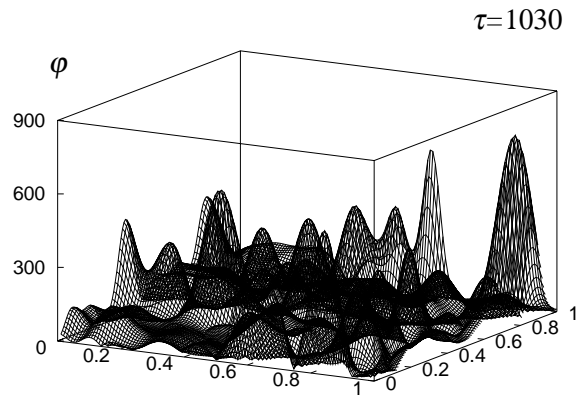}\\[2mm]
\caption[fig3]{\label{phi_1030}
Amplitude of the field with the same range as in
Fig.~\ref{q_1030}.} 
\end{figure}

One surprising fact is that lumps are actually formed just after the
real part of the fluctuations has developed fully even though the
imaginary part has not yet started to grow. But, these lumps have not
yet contained charges inside. This can be gathered in
Figs.~\ref{q_1030} and \ref{phi_1030}, where we respectively show the
charge density and the amplitude of the field at the time just after
the amplitude of the fluctuations in the real direction catches up
with that of the homogeneous mode. Although the field has not settled
down completely, we can already see many lumpy objects, which have no
charge. Notice that they have in fact the right size of the
prospective $Q$ balls: $\sim |K|^{-1/2}m^{-1}$.

\begin{figure}[t!]
\centering
\hspace*{-7mm}
\leavevmode\epsfysize=6.5cm \epsfbox{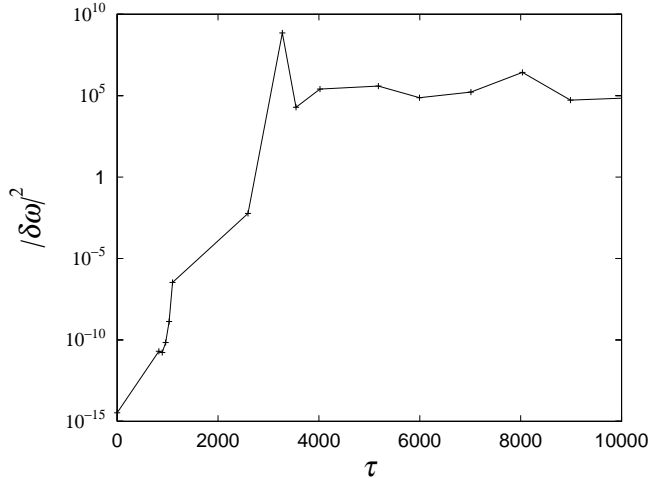}\\[2mm]
\caption[fig3]{\label{omega}
Evolution of the fluctuation of angular velocity. }
\end{figure}

\begin{figure}[t!]
\centering
\hspace*{-7mm}
\leavevmode\epsfysize=6.5cm \epsfbox{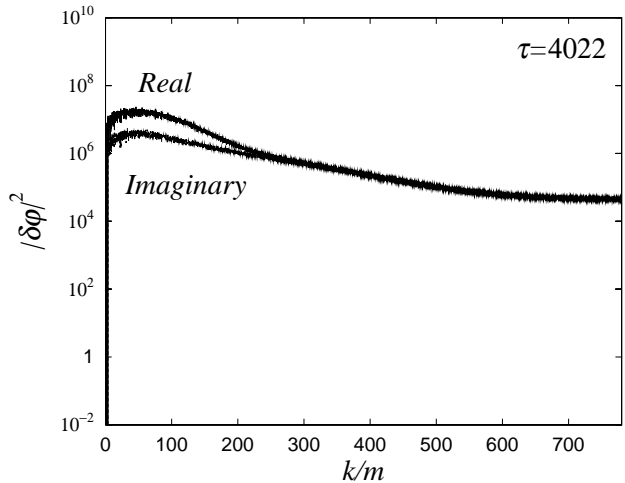}\\[2mm]
\caption[fig3]{\label{sp4022}
Power spectra of fluctuations in the real and imaginary directions
just after the real part fully developed.}
\end{figure}

As time goes on, charges are accumulated into lumps to become $Q$
balls due to mode-mode couplings between fluctuations in the real and
imaginary directions. We show the evolution of
$\langle\delta\omega^2\rangle$ in Fig.~\ref{omega}. It is evaluated
from 
\begin{equation}
    \langle \delta\omega^2 \rangle =
       \frac{\left[(\varphi_2^2 - \varphi_1^2)\varphi_2'
                  + 2\varphi_1\varphi_2\varphi_1' \right]^2}
            {(\varphi_1^2 + \varphi_2^2)^4}
         \langle \delta\varphi_1^2 \rangle
     + \frac{\left[(\varphi_2^2 - \varphi_1^2)\varphi_1'
                  - 2\varphi_1\varphi_2\varphi_2' \right]^2}
            {(\varphi_1^2 + \varphi_2^2)^4}
         \langle \delta\varphi_2^2 \rangle
     + \frac{\varphi_2^2}{(\varphi_1^2 + \varphi_2^2)^2}
         \langle \delta\varphi_1'^2 \rangle
     + \frac{\varphi_1^2}{(\varphi_1^2 + \varphi_2^2)^2}
         \langle \delta\varphi_2'^2 \rangle.
\end{equation}
We see that $\langle\delta\omega^2\rangle$ starts to grow right after
the field fluctuations in the real part have fully developed. We may
regard it as a local evolution of the charge distribution, or, in
other words, the charge accumulation into lumps, which leads to the
$Q$-ball formation. Once it is saturated (at $\tau \sim 3500$), lumpy 
objects will not disintegrate but remain as $Q$ balls because of the
charge conservation. We can also see in Fig.~\ref{sp4022}, where the
spectra of fluctuations in real and imaginary parts are shown. At 
around this time, the amplitude of the imaginary part has almost
matched with the real partner, which also implies the actual 
$Q$-ball formation.

\begin{figure}[t!]
\centering
\hspace*{-7mm}
\leavevmode\epsfysize=6.5cm \epsfbox{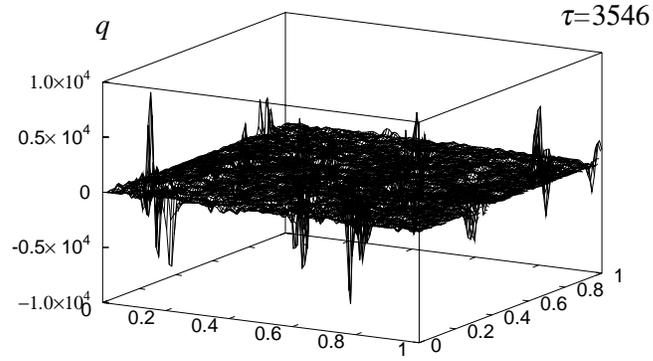}\\[2mm]
\caption[fig3]{\label{q_3546}
Charge density distribution in a small sub-lattice.}
\end{figure}

\begin{figure}[t!]
\centering
\hspace*{-7mm}
\leavevmode\epsfysize=6.5cm \epsfbox{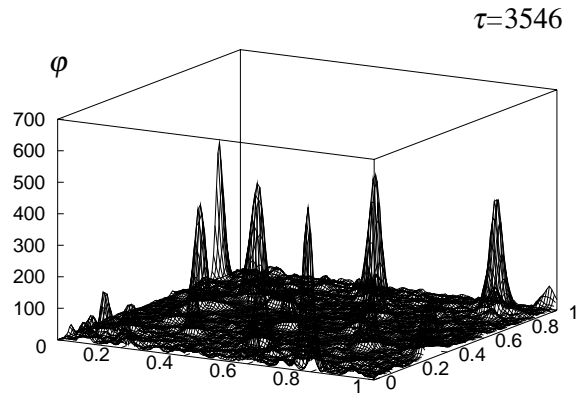}\\[2mm]
\caption[fig3]{\label{phi_3546}
Amplitude of the field in the same region as in
Fig.~\ref{q_3546}.} 
\end{figure}

At this time, the field values settle down, and $Q$ balls and anti-$Q$
balls are completely isolated from the background. Figures
\ref{q_3546} and \ref{phi_3546} show the charge density and the
amplitude of the field after the imaginary part of the field
fluctuation has fully developed. Now we can see ``mature'' $Q$ balls,
and charges of the $Q$ balls conserve from that time on.

\begin{figure}[t!]
\centering
\hspace*{-7mm}
\leavevmode\epsfysize=6.5cm \epsfbox{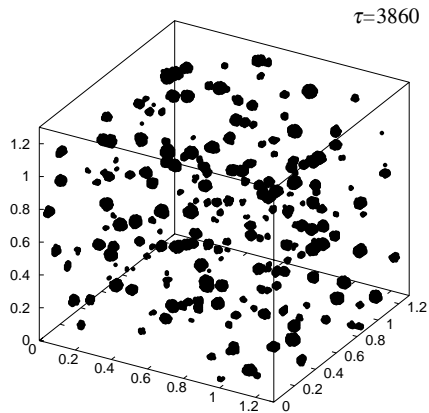}\\[2mm]
\caption[fig3]{\label{3D}
$Q$ balls in three-dimensional lattices with $N=128$ and 
$\Delta \xi=0.01$ at $\tau=3860$.}
\end{figure}

Of course, a $Q$-ball formation is seen on the three-dimensional
lattices. We only show how it appeared in Fig.~\ref{3D}. Here, we plot
the configuration at $\tau=3860$, just after $Q$ balls are completely
formed. The typical size and charge of the $Q$ balls are 
$R_{phys}/m \sim 0.05 \times a(3860) \sim 10$, which almost coincides
with the analytically estimated values, and 
$Q \sim$ (a few) $\times 10^6$

Therefore, it appears that in running mass inflation $Q$ balls are
produced not only by the negative pressure which makes the homogeneous
condensate to fluctuate at the same scale as the $Q$-ball size, but
the mode-mode coupling plays a crucial role in creating locally
distributed charge density. In this sense we can state that in a
potential such as Eq.(\ref{qpot}) a kind of preheating leads to
$Q$-ball formation.

\section{Reheating through evaporation}

So far, we have seen it is very natural for the inflaton condensate to
deform into $Q$ balls in a certain inflaton potential such as
Eq.(\ref{qpot}). As a consequence, it may provide the alternative way
of reheating after inflation, namely reheating through evaporation,
which leads to slower reheating process than usually considered, as
we claimed in Ref.~\cite{EKM}. $Q$ balls in the potential (\ref{qpot})
are unstable for decaying into particles lighter than $\sim m$
\cite{em}. If the decay products are pairs of fermions, the decay
process occurs only on the surface of $Q$ balls because of the Pauli
blocking inside the $Q$ ball \cite{Cohen}.

A $Q$ ball of size $R \sim |K|^{-1/2}m^{-1}$ forms when the
fluctuations grow nonlinear. Since the growth rate of fluctuation is 
$\sim |K|m$, the Hubble parameter at the formation time will be
estimated as
\begin{equation}
    H_{f} \sim \gamma |K|m,
\end{equation}
where $\gamma \sim 2\times 10^{-2}$ is some numerical factor derived
from numerical simulations. As we assume that $|K|\ll 1$, we can
approximate the decrease in the amplitude of the oscillations by 
$\phi_f \sim \phi_i(H_f/H_i)$ as in the matter dominated era, where
$\phi_i \simeq M_{\rm P}/\sqrt{12\pi}$, denotes the amplitude at the
end of inflation in chaotic model, and $H_i\sim m$ when the
oscillations begin. Therefore, the total charge of a $Q$ ball is given
by 
\begin{equation}
    Q \sim \frac{4\pi}{3} R^3 n_q 
      \sim \frac{1}{9}\beta \zeta^2 \gamma^2 |K|^2 R^3 m M_{\rm P}^2, 
\end{equation}
where $n_q = \beta \omega\phi_0^2$, $\phi_0 \simeq \zeta \phi_f$, and
$\beta \ll 1$ and $\zeta \gtrsim 1$ are numerical factors.

As we shall see below, the evaporating reheat takes over the usual
perturbative decay in relatively large coupling where $h\lesssim 1$. 
In this regime, the evaporation rate is saturated by
\begin{equation}
    \label{evap}
    \frac{dQ}{dt} \lesssim \frac{\omega^3 A}{192\pi^2}\,,
\end{equation}
where $A\simeq 4\pi R^2$ is the surface area of the $Q$ ball, and 
$\omega \simeq m$. Thus, we obtain the decay rate of the $Q$ ball
through evaporation as 
\begin{equation}
    \label{evap2}
    \Gamma_Q = \frac{1}{Q}\frac{dQ}{dt}
        \simeq \frac{3}{16\pi \beta\zeta^2\gamma^2|K|^{3/2}}
                \left(\frac{m}{M_{\rm P}}\right)^2 m\,.
\end{equation}
Note that the decay rate is determined by the ratio 
$m/M_{\rm P} \simeq 10^{-6}$, which is fixed by the anisotropies
seen in the cosmic microwave background radiation. Therefore, even
though we are in relatively large coupling limit, the decay rate
mimics that of a Planck suppressed interaction of the inflatonic 
$Q$ ball to the matter fields.

Now let us compare the rate (\ref{evap2}) with the perturbative decay
rate $\Gamma_{pert} = h^2m/8\pi$. Taking the ratio, we have
\begin{equation}
    \label{ratio}
    \frac{\Gamma_Q}{\Gamma_{pert}} \simeq 
        \frac{3}{2\beta\zeta^2\gamma^2|K|^{3/2}h^2}
          \left(\frac{m}{M_P}\right)^2.
\end{equation}
When this ratio is less than unity, the evaporation process from the
$Q$-ball surface suppresses the reheating. This is achieved if 
$h \gtrsim 10^{-3}$ for reasonable parameters, such as $|K| \sim 0.1$,
$\beta \sim 10^{-2}$, $\gamma \sim 2\times10^{-2}$, and 
$\zeta \sim 2$. These values of the parameters are derived from
numerical simulations on the three-dimensional lattices: 
$Q \sim 3 \times 10^6$, for example. Since the reheating temperature
can be estimated as
\begin{equation}
    T_{RH} \simeq 0.1 \sqrt{\Gamma M_P} \propto \Gamma^{1/2},
\end{equation}
it will be suppressed by a factor $\sim 2\times 10^{-3}$ for 
$h \sim 1$.

In general $K$ and $h$ are not independent quantities but are related
to each other by $|K| \sim C h^2/16\pi^2$, where $C$ is effective
number of bosonic and fermionic loops. If the inflaton sector does not
belong to a hidden sector, it is very natural that the inflaton
coupling to other matter fields is relatively large, i.e. 
$h\gtrsim (m/M_{\rm P})$. We can thus have easily a situation where
$|K| \sim 0.1$, if, e.g., $C \sim 10$ and $h \sim 1$.

To end this section, we will comment on (fermionic) preheating 
\cite{heitmann,Kofman}. In the preheating stage, it is very easy to
transport inflaton energy into a pair of massless fermions in an
efficient way with the maximum physical momentum
\begin{equation}
    \label{mode1}
    k_{max} \propto (h\phi m)^{1/2}\,,
\end{equation}
where $\phi$ is the amplitude of the oscillating inflaton.
Even though fermionic preheating is efficient, the whole inflaton
energy is not transferred in this process. The energy density stored
in the fermions remains small compared to the inflaton energy density,
i.e. the comoving number density and the energy density of the
massless fermions follow 
\begin{equation}
    n_f \sim \frac{k_{max}^3}{6\pi^2} 
        \sim \frac{(h\phi m)^{3/2}}{6\pi^2}, \qquad
    \rho_f \sim \frac{k_{max}^4}{8\pi^2} 
           \sim \frac{h^2\phi^2m^2}{8\pi^2}. 
\end{equation}
The ratio of fermion energy density compared to the inflaton energy
density after the end of fermion preheating behaves as
\begin{equation}
    \frac{\rho_{f}}{\rho} \sim \frac{h^2}{8\pi^2} \ll 1\,.
\end{equation}
Hence most of the energy density of the inflaton does not transfer to
that of the fermions. This result is quite robust and remains valid
even if the fermions have a non-zero bare mass \cite{giudice}.

Even if fermions are produced, they cannot scatter inflaton
quanta off its condensate \cite{heitmann}, unlike in the case of
bosonic preheating with an interaction $g^2\phi^2\chi^2$. The zero
mode of the inflaton thus remains intact as a coherent condensate
which will fragment into $Q$ balls later. Notice that, as mentioned
earlier, a $Q$-ball formation is reminiscent of bosonic preheating due
to the presence of attractive self coupling of the inflaton, which
stems in the logarithmic term in the potential. Therefore, $Q$-ball
formation appears to be a robust feature of any inflationary model
with a potential allowing for a $Q$-ball solution. Moreover, for
a flatter potential, the size of the $Q$ ball is correlated with
the charge, as in the context of the Affleck-Dine condensate with the 
gauge-mediated SUSY breaking. This means that the suppression on the
reheating will be much more efficient in those cases.

\section{Conclusion and Discussion}

Inflaton condensate fragmentation and the formation of lumps may be a
rather generic, albeit in many cases a transient feature of
preheating. However, for the running mass inflaton with $K<0$, and
possibly for other types of inflation admitting at least temporarily a
mass term growing slower than $\vert \Phi\vert^2$, the final outcome
is markedly different from the usual preheating. The effective
equation of state, averaged over field oscillation in the real
direction, has negative pressure, which is the cause of the condensate
break-up.  If the inflaton happens to be a complex field so that it
admits soliton solutions carrying a global charge, one would typically
expect an inflatonic $Q$ and anti-$Q$ ball pair to be generated
when within a $Q$-ball forming region the field fluctuations in both
the real and the imaginary directions exceed the coherent background
field value. 

To verify the generation of inflatonic solitons it is not sufficient
to follow only the evolution of the field amplitudes. In the real
direction the fluctuations grow very fast to the size of the 
prospective soliton lumps, but this by itself is not sufficient to
ensure $Q$-ball formation. Indeed, field fluctuations in the imaginary
direction do not grow at all in the linear approximations. To see what
really happens, one has to perform simulations in a configuration
space on the lattices, as we have done in the present paper. 

In our simulations we can see the formation of lumpy objects but
already at relatively early stages when the mean fluctuations in the
imaginary direction have not yet develope. These objects have not
become $Q$ balls yet, but they eventually accumulate charges to become
$Q$ and anti-$Q$ ball pairs. We believe that the reason is that there
will be large local fluctuations which, once pushed over the limit
after which bound solitonic states can exist, cannot fluctuate
back. The attractive inflaton self-interaction generates locally
distributed $U(1)$ charges through mode-mode couplings which then
evolve into semi-stable inflatonic $Q$ balls. The formation of such
solitonic states is non-trivial and stands in contrast to the lumps of
the $K>0$ case; the latter is likely to be an evanescent feature that
would merely disintegrate and decay away very fast.

Once produced, these $Q$ balls will decay by evaporation and hence
give rise to a ``slow'' reheating because of the suppression which
comes from the surface-volume ratio. In particular, this effect
becomes more efficient for rather larger coupling $h \sim 1$, which
might help the disastrous situations such as the gravitino problem in
the supersymmetric scenarios. In fact, the surface evaporation through
$Q$ ball is perhaps the most natural way to obtain low reheat
temperatures inspite of having order one inflaton coupling to the
fermions, which solves not only the gravitino over-abundance problem
but also moduli, and dilaton, especially when the inflationary scale
is sufficiently large.

Although we have looked upon the running-mass potential, existence of
the $Q$-ball solution is a very generic feature in a certain class of
models. It would be interesting to study whether such inflatonic $Q$
balls could be seen to arise also in other inflationary models such 
as hybrid inflation, new inflation, etc., and study in these
directions are in progress.

\section*{Acknowledgement}
A.M. is thankful to Rouzbeh Allahverdi, Mar Bastero-Gil and Altug
Ozpineci for helpful discussion, A.M. acknowledges the support of
{\it The Early Universe network} HPRN-CT-2000-00152, and a kind
hospitality of the Helsinki Institute of Physics where this work has
been carried out. K.E. acknowledges the Academy of Finland
grant 51433. Part of the numerical calculations was carried out
on VPP5000 at the Astronomical Data Analysis Center of the National
Astronomical Observatory, Japan.


\begin{references}

\bibitem{pert} 
  A.D. Dolgov and A.D. Linde, Phys. Lett. {\bf B116},329 (1982); \\
  L. F. Abbott, E. Farhi, and M.B. Wise, 
  Phys. Lett. {\bf B117}, 29 (1982).

\bibitem{traschen}
  J. Traschen and R. H. Brandenberger,
  Phys. Rev. D {\bf 42}, 2491 (1990); \\
  L. A. Kofman, A. D. Linde, and A. A. Starobinsky,
  Phys. Rev. Lett. {\bf 73}, 3195 (1994).

\bibitem{gravitino}
  M. Y. Khlopov and A. D. Linde, 
  Phys. Lett. {\bf B138}, 265 (1984);\\
  J. R. Ellis, J. E. Kim, and D. V. Nanopoulos,
  Phys. Lett. {\bf B145}, 181 (1984); \\
  M. Kawasaki and T. Moroi,
  Prog. Theor. Phys. {\bf 93}, 879 (1995);\\
  J. Ellis, D.V. Nanopoulos, K.A. Olive and S-J. Rey, Astropart. Phys.
  {\bf 4}, 371 (1996);\\
  for a recent calculation see: 
  M. Boltz, A. Brandenburg and W. Buchmuller, Nucl. Phys. {\bf B606}, 518 
  (2001);\\
  for non-perturbative production of gravitino see:
  A. L. Maroto and A. Mazumdar, Phys. Rev. Lett {\bf 84}, 1655 (2000);\\
  R. Kallosh, L. Kofman, A. Linde and A. Von Proeyen, Phys. Rev D {\bf 61}, 
  103503 (2000);\\
  R. Allahverdi, M. Bastero-Gil, and A. Mazumdar, Phys. Rev. D. {\bf 64},
  023516 (2001).\\
  H.P. Nilles, M. Peloso and L. Sorbo, JHEP 0104, 004 (2001).



\bibitem{EKM}
  K. Enqvist, S. Kasuya, and A. Mazumdar, hep-ph/0204270.

\bibitem{coleman}
  S. Coleman, Nucl. Phys. B {\bf 262}, 263 (1985); \\
  T. D. Lee and Y. Pang, Phys. Rept. {\bf 221}, 251 (1982).



\bibitem{copeland} 
  G. N. Felder, J. Garcia-Bellido, P. B. Greene, L. Kofman, A. D. Linde, 
  and I. Tkachev, Phys. Rev. Lett. {\bf 87}, 011601 (2001);\\
  G. N. Felder, L. A. Kofman and A.D. Linde, Phys. Rev. D {\bf 64}, 123517
  (2001);\\ 
  E.J. Copeland, S. Pascoli, and A. Rajantie, Phys. Rev. {\bf D65}, 
  103517 (2002). 
  
\bibitem{john} J. McDonald, hep-ph/0105235.

\bibitem{AD} I.A. Affleck and M. Dine, Nucl. Phys. {B249}, 361 (1985).

\bibitem{DKS}
  G. Dvali, A. Kusenko, and M. Shaposhnikov, 
  Phys. Lett. B {\bf 417}, 99 (1998).

\bibitem{KuSh}
  A. Kusenko and M. Shaposhnikov, Phys. Lett. B {\bf 418}, 46 (1998).

\bibitem{KK1}
  S. Kasuya and M. Kawasaki, Phys. Rev. D {\bf 61}, 041301 (2000).

\bibitem{KK4}
  S. Kasuya and M. Kawasaki, Phys. Rev. Lett. {\bf 85}, 2677 (2000).

\bibitem{KK3}
  S. Kasuya and M. Kawasaki, Phys. Rev. D {\bf 64}, 123515 (2001).


\bibitem{empl} 
  K. Enqvist and J. McDonald, Phys. Lett. {\bf B418}, 46 (1998).

\bibitem{em}
  K. Enqvist and J. McDonald, Nucl. Phys. {\bf B538}, 321 (1999).

\bibitem{KK2}
  S. Kasuya and M. Kawasaki, Phys. Rev. D {\bf 62}, 023512 (2000).

\bibitem{LyRi} 
  D. Lyth and A. Riotto, Phys. Rep. {\bf 314}, 1 (1999).

\bibitem{Cohen}
  A. Cohen, S. Coleman, H. Georgi, and A. Manohar,
  Nucl. Phys. B {\bf 272}, 301 (1986).

\bibitem{turnerpressure} 
  M.S. Turner, Phys. Rev. {\bf D28}, 1243 (1983).

\bibitem{johnpressure} 
  J. McDonald, Phys. Rev. {\bf D48}, 2573 (1993).

\bibitem{asko} A. Jokinen, hep-ph/0204086.

\bibitem{MuVi}
  T. Multam\"{a}ki and I. Vilja, Phys. Lett. B {\bf 535}, 170 (2002).
\bibitem{EnJoMuVi}
  K. Enqvist, A. Jokinen, T. Multam\"{a}ki, and I. Vilja,
  Phys. Rev. D {\bf 63}, 083501 (2001).

\bibitem{Kasuya1}
  S. Kasuya, 
  in: {\it Proceedings Of The 7th International Symposium On Particles,
  Strings And Cosmology (PASCOS-99)}, K. Cheung et al., 301 (1999).

\bibitem{Kasuya2}
  S. Kasuya, Phys. Lett. B {\bf 515}, 121 (2000).

\bibitem{heitmann}
  J. Baacke, K. Heitmann, and C. Patzold,
  Phys. Rev. D {\bf 58}, 120513 (1998).

\bibitem{Kofman}
  P. Greene and L. Kofman, Phys. Lett. B {\bf 448}, 6 (1999).

\bibitem{giudice}
  G.F. Giudice, M. Peloso, A. Riotto, and I. Tkachev,
  JHEP {\bf 08}, 014 (1999). 


\end{references}
\end{document}